Localized vorticity enhancement through superhelical coherent structure in observed tornadic supercells.
Marcus Büker, Luke Odell, UW-Madison; Joshua Wurman, CSWR; Jim Marquis, PSU. To be submitted to Geophysical Research Letters.

## ABSTRACT

New diagnostic methods are presented for localized, barotropic vorticity evolution in tornadic environments. These methods focus on superhelicity, a quantity shown to be strongly related to local maxima in vorticity tendency. Mobile Doppler radar retrievals from three well-known cases of tornadogenesis were studied with this new approach. The results show promise for detecting and tracking coherent dynamical features known to be related to tornadogenesis, as well as detecting signals of imminent tornadogenesis at an earlier stage than vorticity tendency analysis. Furthermore, these methods may provide physical insight into vortex-vortex interactions and vortex modes in the vicinity of the mesocyclone.

**SECTION I. INTRODUCTION**

Vorticity production associated with the downdraft region of a supercell thunderstorm has been repeatedly linked to tornadogenesis in numerous studies [e.g. *Fujita,* 1975; *Barnes,* 1978; *Lemon and Doswell,* 1979; *Markowski et al.,* 2003]. Vorticity in this region can be generated by both barotropic and baroclinic processes [e.g. *Davies-Jones*, 2001; 2006]. It is still unclear which mechanism (or a combination thereof) primarily instigates tornado formation. Recently, *Davies-Jones* [2008; hereafter *DJ08*] investigated how a strong maximum in near-surface vertical vorticity could be barotropically produced by transport of angular momentum down to the surface from aloft by a descending rain curtain. It has been shown that downdraft-induced (both barotropically and baroclinically-generated) horizontal vorticity can be barotropically tilted into the vertical through differential vertical motion near the low-level mesocyclone, where the updraft and downdraft interact [e.g. *Straka, et al.*, 2007; *Markowski et al.*, 2008]. Recent analyses of Dual-Doppler fields show significant barotropic contributions to the vorticity [e.g. *Wurman, et al.*, 2007, *Marquis et al.*, 2008], which are typically diagnosed as the tilting and stretching terms in the (barotropic, or BT) vertical vorticity tendency equation:

$$\dot\zeta_{BT} = \boldsymbol{\omega}\cdot\nabla w = \underbrace{\omega_1 w_x + \omega_2 w_y}_{tilting} + \underbrace{\omega_3\, w_z}_{stretching}$$

(1)

Using this equation as a diagnostic tool, these terms show general regions where vertical vorticity will be changing with time. However, in many cases, regions of significant vorticity tendency are broadly distributed along quasi-linear boundaries (e.g. gust fronts, vortex sheets, and vortex rolls), making it difficult to predict where a local maximum in vorticity (such as a tornado) may form. Diagnosing the local maxima of vorticity tendency can more precisely identify regions of rapid vortex growth, given some level of continuity in the forcing distribution.

This letter will be set out in the following manner. Firstly, we demonstrate the utility of examining local extrema of vorticity tendency in the context of tornado dynamics using three-dimensional flow fields from Doppler-on-Wheels (DOW) mobile radar retrievals [*Wurman et al.,* 1997; 1999; 2000]. Secondly, we show how the distribution of a Galilean-invariant scalar (superhelicity density) is related to these maxima. Finally, we illustrate how these diagnostics show a consistent spatiotemporal pattern (and possible periodicity) in the crucial timeframe leading up to tornadogenesis.

**SECTION II. THEORY**

Local maxima in vortex intensification through a barotropic process are governed by the vector Laplacian of the vortex stretching term. Given a velocity field, **u**, and a vorticity field, $\boldsymbol{\omega} = \nabla \times \mathbf{u}$, in an inviscid, barotropic fluid, the evolution of vorticity reduces to

$$\frac{d\boldsymbol{\omega}}{dt} = (\boldsymbol{\omega}\cdot\nabla)\mathbf{u} = \mathbf{T}$$

(2)

Taking the negative Laplacian of both sides yields

$$-\nabla^2\left(\frac{d\boldsymbol{\omega}}{dt}\right) = -\nabla^2\left((\boldsymbol{\omega}\cdot\nabla)\mathbf{u}\right) = -\nabla\left(\nabla\cdot\left((\boldsymbol{\omega}\cdot\nabla)\mathbf{u}\right)\right) + \nabla\times\nabla\times(\boldsymbol{\omega}\cdot\nabla)\mathbf{u} = -\nabla\left(\nabla\cdot\mathbf{T}_\chi\right) + \nabla\times\nabla\times\mathbf{T}_\varphi.$$

The vortex stretching term, $\mathbf{T}$, has been decomposed into irrotational and non-divergent components

$$(\boldsymbol{\omega}\cdot\nabla)\mathbf{u} = \mathbf{T}_\chi + \mathbf{T}_\varphi, \text{ so } \nabla\cdot\mathbf{T}_\varphi = |\nabla\times\mathbf{T}_\chi| = 0.$$

Assuming $\nabla\cdot\mathbf{u} = 0$, and using Green's second vector identity [*Fernández-Guasti*, 2012]

$$\nabla^2(\mathbf{P}\cdot\mathbf{Q}) = \mathbf{P}\cdot\nabla^2\mathbf{Q} - \mathbf{Q}\cdot\nabla^2\mathbf{P} + 2\nabla\cdot\left[(\mathbf{Q}\cdot\nabla)\mathbf{P} + \mathbf{Q}\times\nabla\times\mathbf{P}\right],$$

while substituting $\mathbf{P} = \mathbf{u}$ and $\mathbf{Q} = \boldsymbol{\omega}$, rearranging, and taking the negative gradient, yields

$$-\nabla^2\mathbf{T} = \frac{1}{2}\nabla\left[S + \mathbf{u}\cdot\nabla^2\boldsymbol{\omega} - \nabla^2 H\right] + \nabla\times\nabla\times\mathbf{T}_\varphi$$

(3)

where $S \equiv \boldsymbol{\omega}\cdot\nabla\times\boldsymbol{\omega}$ is the local *superhelicity* density [*Hide*, 1989; 2002] and $H \equiv \mathbf{u}\cdot\nabla\times\mathbf{u} = \mathbf{u}\cdot\boldsymbol{\omega}$ is the local helicity density. (For brevity, we will hereafter refer to these local quantities as *superhelicity* and *helicity*.) Thus, regions of local vortex intensification can be diagnosed, at least in part, by the distribution (i.e. gradient) of superhelicity.

While superhelicity has been used chiefly in the turbulence literature in the context of helicity evolution [e.g. *Ricca*, 1994; *Galanti and Tsinober*, 2006; *Jacobitz et al.*, 2011], it has only been referenced on occasion with regard to supercell dynamics. *Kanak and Lilly* [1996] examined the quantity in terms of ambient vertical wind shear, showing that, unlike helicity, superhelicity is Galilean invariant and is "very sensitive to loops in the hodograph". *Shiqiang and Zhemin* [2001] is, to our knowledge, a unique example of a study explicitly using superhelicity as a diagnostic in a supercell simulation. They hypothesized using superhelicity as a marker for maturity of supercell development, as they found high negative superhelicity values at low levels in the storm during its developing stages, while positive values steadily increased later in the storm lifecycle. Here, we develop a more intrinsic and quantitative diagnostic role for superhelicity.

With some simplifying assumptions, the relationship between superhelicity and local vortex intensification can be understood in the context of the vortex interaction between a local downdraft surge and a mesocyclone (Figure 1). In the region of the mid-level mesocyclone, we assume a locally homogenous region of pure vertical vorticity, $\bar{\boldsymbol{\omega}} = \nabla\times\bar{\mathbf{u}}_\mathbf{h}$, where $\bar{\mathbf{u}}_\mathbf{h} = \bar{u}\hat{\mathbf{i}} + \bar{v}\hat{\mathbf{j}}$ and $\hat{\mathbf{k}}\cdot\bar{\boldsymbol{\omega}} = \bar{\omega}_3 = \bar{v}_x - \bar{u}_y$. Now, assume a segment of pure horizontal vorticity $\boldsymbol{\omega}'_h$ moves into the region, where $\boldsymbol{\omega}'_h = \nabla\times\mathbf{u}'$, and $\mathbf{u}' = u'\hat{\mathbf{i}} + v'\hat{\mathbf{j}} + w'\hat{\mathbf{k}}$. Given a (non-zero) horizontal gradient of this vorticity, then (if $\nabla\cdot\mathbf{u}' = 0$),

$$|\nabla\times\boldsymbol{\omega}'_h| = \hat{\mathbf{k}}\cdot(\nabla\times\boldsymbol{\omega}'_h) = u'_{xz} + v'_{yz} - w'_{xx} - w'_{yy} = -\nabla^2 w'.$$

Thus the curl of the horizontal vorticity is akin to a local maximum in vertical motion (i.e. updraft or downdraft pulse). Superhelicity in this circumstance will be $S = \bar{\boldsymbol{\omega}}\cdot\nabla\times\boldsymbol{\omega}' = -\bar{\omega}_3\nabla^2 w'$, with the vertical gradient,

(4)
$$S_z = -\bar{\omega}_3\nabla^2 w'_z = -\nabla^2(\bar{\omega}_3 w'_z)$$

representing the local maxima in pure vertical vortex stretching. Referring to Figure 1, the circulation associated with the downdraft pulse results in a ring of horizontal vorticity ($\boldsymbol{\omega}_h$). This causes a differential advection of vortex lines associated with the mesocyclone. Horizontal gradients of ($\boldsymbol{\omega}_h$) create regions of superhelicity ($S$); vertical gradients of $S$ will also exist given the local nature of the downdraft pulse. Local maxima and minima of vorticity tendency will then occur in planes above and below the region of maximum downdraft. This particular configuration results in a dipole pattern of positive (negative) vortex stretching below (above) the region of maximal horizontal vorticity gradient. In these regions, a vertical gradient of superhelicity exists. In the event of a downward propagating horizontal vortex feature (such as a localized downdraft pulse), the resultant evolution of vorticity would be similar to axisymmetric results shown in *DJ08*, where a downdraft pulse (associated with rain-induced drag) was found to concentrate vorticity to smaller radii near the surface, with vorticity more dispersed aloft (see, for example, contours of angular momentum in Figure 6 of *DJ08*). We hypothesize that this mechanism can be responsible for the downward development of an intense misocyclone on the periphery of a mesocyclone.

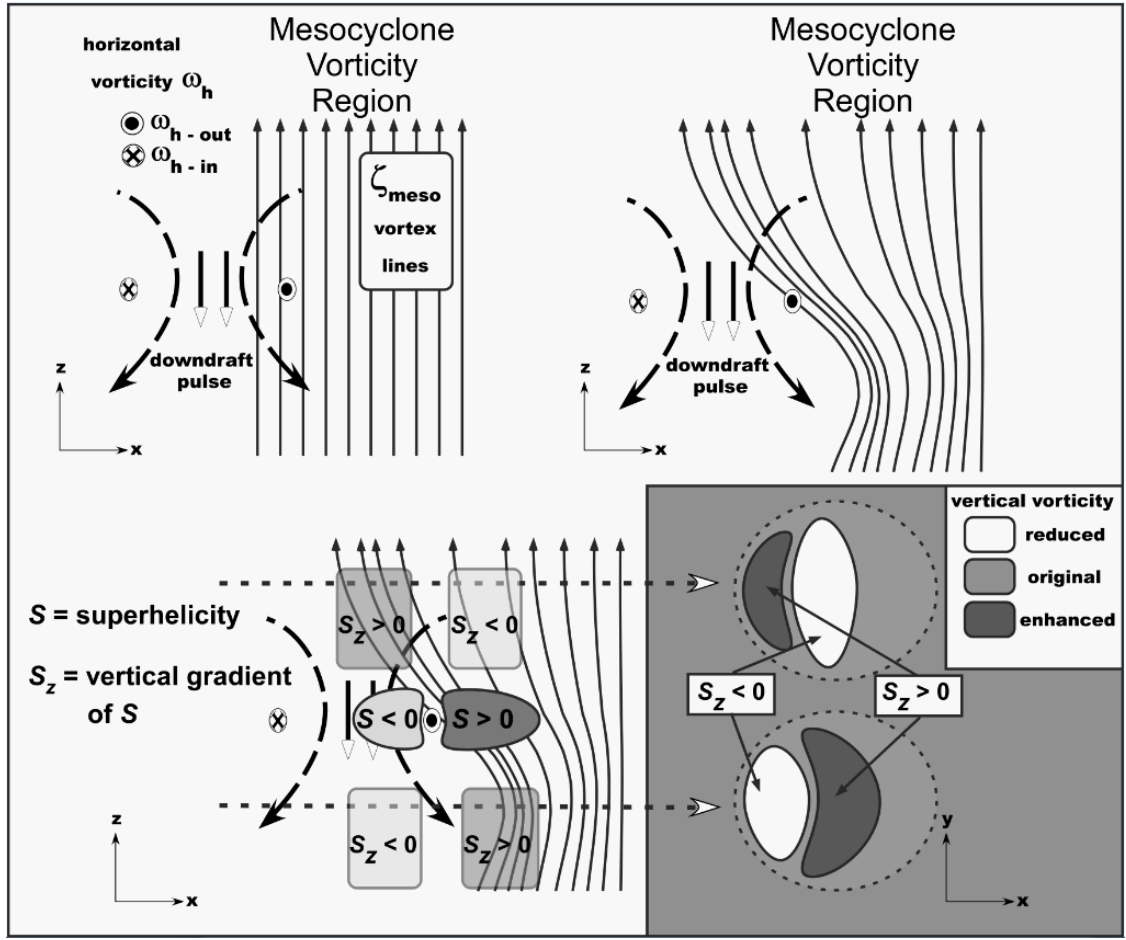

Figure 1. Schematic diagram of superhelical interaction between a flanking downdraft pulse and mesocyclone vorticity.

## SECTION III. DATA AND METHODOLOGY

Three-dimensional DOW retrievals of kinematic fields (3D components of wind and vorticity, including stretching and tilting contributions to vertical vorticity) were obtained from the Center for Severe Weather Research (CSWR) for three tornadic supercell cases (for details regarding dual-Doppler analysis methods, see appendix in *Kosiba et al.*, [2013]). Geographic locations, grid resolution, vertical extent of gridded data, analysis times, and approximate grid-relative mesocyclone locations are given in Table 1.

### A. Cases studied

*1) Goshen*

This premiere observational case captured the full lifecycle of a strong tornadic supercell that occurred near Goshen, Wyoming on 05 June 2009. Observations and analyses are documented in several recent papers (e.g. *Markowski et al.* [2012a,b; hereafter *MK12*]; *Kosiba et al.* [2013; hereafter *K13*]). Two DOW units (DOW6 and DOW7) were well-positioned for dual-Doppler volume scans for the period of 2142-2218 UTC, during when tornadogenesis occurred.

Gridded analyses were provided by CSWR for the period 2142-2202 UTC, yielding 11 analysis times. Grid coordinates (*x* - east, *y* - north) are relative to the DOW6 position.

*2) Argonia*

DOW units (DOW2 and DOW3) observed a supercell and weak tornado from mesocyclone formation to early decay near Argonia, Kansas on 05 June 2001 [*Dowell, et al.* 2002]. This was one of four cases analyzed by *Marquis, et al.* [2012, hereafter *MQ12*]. Dual-Doppler retrievals were obtained for the period of 0014-0036 UTC. Gridded kinematic data were provided by Jim Marquis (through CSWR) for the period 0025-0034 UTC, yielding 7 analysis times. In this case, grid coordinates are referenced to an Earth-relative position.

*3) Orleans*

A more spatiotemporally limited case provides data during the mature and early decaying stage of a moderate tornado near Orleans, Nebraska on 22 May 2004 [*Wurman, et al.* 2010; hereafter *W10*]. While single-Doppler observations were available during an earlier, weaker phase of the tornado, the DOW2 and DOW3 units were not positioned for synchronized volume scans until 2259 UTC (well after tornadogenesis), and analyses were only available up to 1 km altitude. Gridded kinematic data were provided by CSWR for the period 2259-2309 UTC, yielding 7 analysis times. The DOW2 position serves at the grid reference in this case.

| **Location** **Date** | **Grid** $\Delta(x,y,z)/$ $z_{max}$ (m) | | | Doppler Retrieval Time Intervals | | | | | | | | | | |
|---|---|---|---|---|---|---|---|---|---|---|---|---|---|---|
| | | | | 1 | 2 | 3 | 4 | 5 | 6 | 7 | 8 | 9 | 10 | 11 |
| **Goshen, WY** 05 Jun 2009 | 100/ 4000 | **Time** (UTC) | | 2142 | 2144 | 2146 | 2148 | 2150 | 2152 | 2154 | 2156 | 2158 | 2200 | 2202 |
| | | Center (km) | x | -15.0 | -14.8 | -14.0 | -13.5 | -13.0 | -11.0 | -10.0 | -9.0 | -8.0 | -6.5 | -5.0 |
| | | | y | 18.6 | 18.5 | 18.4 | 18.0 | 17.5 | 17.0 | 17.0 | 16.5 | 16.0 | 16.0 | 16.0 |
| **Argonia, KS** 05 Jun 2001 | 150/ 3000 | **Time** (UTC) | | 0025 | 0028 | 0029 | 0030 | 0032 | 0033 | 0034 | | | | |
| | | Center (km) | x | 1.3 | 2.3 | 2.8 | 3.4 | 4.0 | 5.0 | 5.8 | | | | |
| | | | y | 1.0 | 1.5 | 0.7 | 0.0 | 0.0 | 0.0 | 0.6 | | | | |
| **Orleans, NE** 22 May 2004 | 100/ 1000 | **Time** (UTC) | | 2259 | 2300 | 2301 | 2303 | 2305 | 2307 | 2309 | | | | |
| | | Center (km) | x | -12.9 | -12.1 | -10.4 | -8.2 | -6.6 | -4.7 | -3.3 | | | | |
| | | | y | 2.5 | 2.4 | 2.2 | 2.0 | 1.6 | 1.1 | 0.3 | | | | |

Table 1. Dual-Doppler retrieval data times for three tornadic cases. Mesocylone circulation positions (earth or grid relative) are given for each retrieval time.

**B. Peak-value time-height analysis and cross-correlations**

Using four-dimensional (*x,y,z,t*) gridded data from the three cases, we calculated four key quantities: TOT, the total barotropic contribution to vertical vorticity tendency (the sum of the stretching and tilting terms); LTOT, the negative Laplacian of TOT; *S*, the superhelicity, and $S_z$, the vertical gradient of superhelicity. Using these four quantities, along with vertical vorticity and velocity, a time-height analysis of peak values proximal to the mesocyclone location was constructed. Peak proximal values at each time and height were computed by searching a 2 km radius around the mesocyclone position (see Table 1), and averaging the highest 10 grid point values for each quantity. The choice of radius was based on the desire to include the progression of dynamical features along the periphery of the mesocyclone as well as features near the gust front. Averaging over 10 points captured the top 1-2% of values in the target region, while providing a buffer against spuriously large values. In cases where there was not enough data within the circle at a certain level or time for a valid sample size, that level was omitted from the variable time-height grid. In most cases, this occurred near the top and bottom of the domain, where finite difference calculations involving multiple vertical levels resulted in the lowest and highest 1-3 vertical levels (corresponding to the lowest and highest 100-300 meters) being discarded. As seen in the next section, a high level of coherency (as well as consistency with other studies) in various dynamical features was observed using this method.

Additionally, cross-correlations were computed to determine relative strength of spatiotemporal relationship between these variables. Another term, UDZ = $\partial/\partial z \left( \mathbf{u} \cdot \nabla^2 \boldsymbol{\omega} \right)$, was included, as it was the only other term on the RHS of (3) that had typical magnitudes comparable to the superhelicity gradient.

It should be noted that there are likely some regions of error in the retrieval data at higher altitudes (above 2 km). This is especially true for variables depending on vertical velocity (see, for example, the discussion in *MQ12*). However, our time-height analysis reveals strong vertical continuity in the fields, even at higher altitudes.

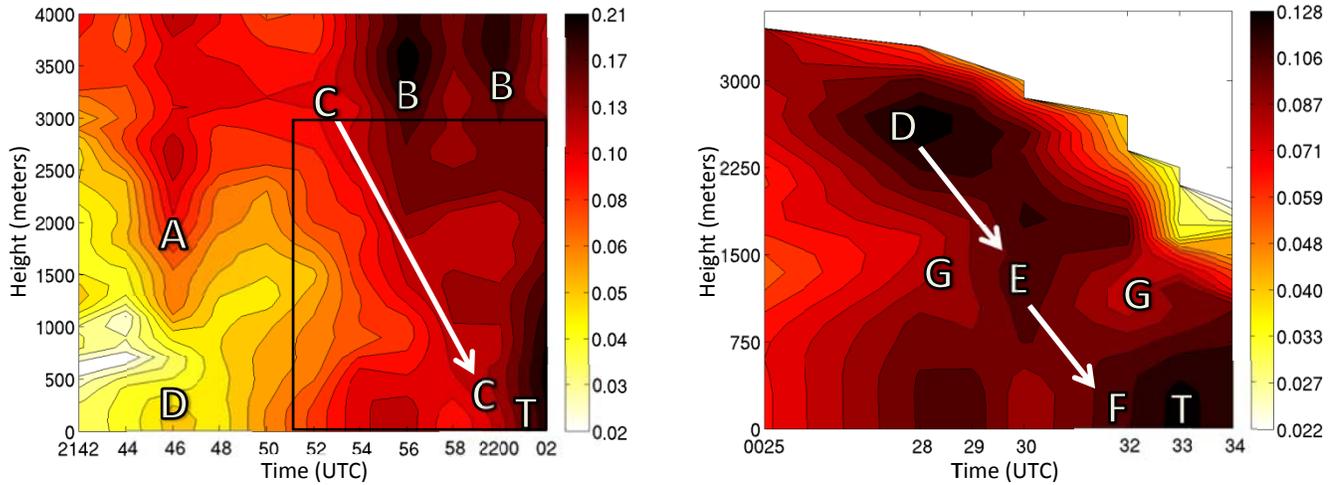

Figure 2. Time-height analysis of peak vorticity values (s$^{-1}$) for dual-Doppler retrievals in the Goshen case (left) and Argonia case (right). "T" denotes observed time of tornadogenesis. (*Please note that all figures are contoured on a $log_{10}$ scale.*)

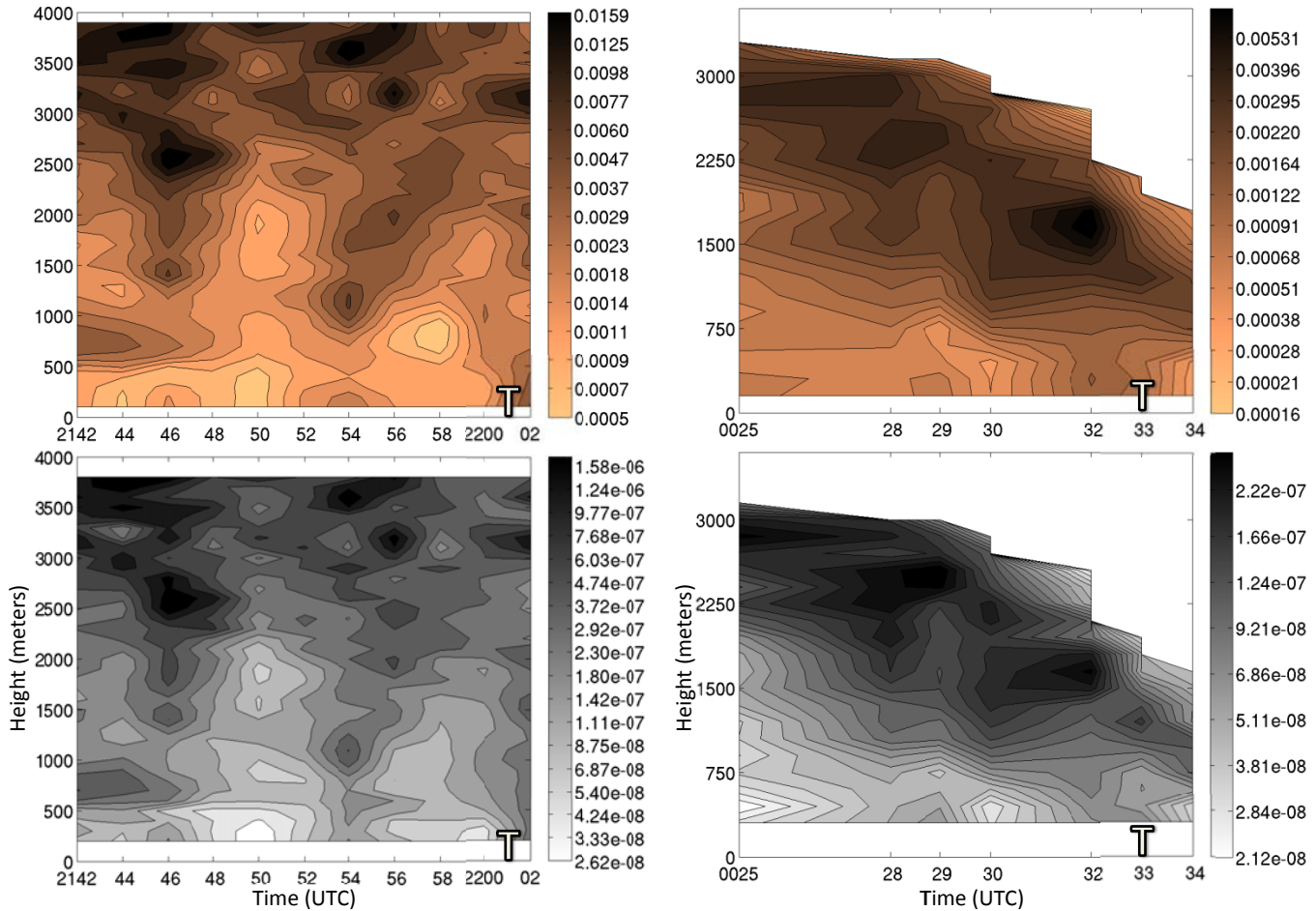

Figure 3. As in Figure 2, but with time-height analysis of peak vertical vorticity tendency (TOT, s$^{-2}$; top row) and negative Laplacian of this quantity (LTOT, m$^{-2}$s$^{-2}$; bottom row) for the Goshen case (left column) and Argonia case (right column).

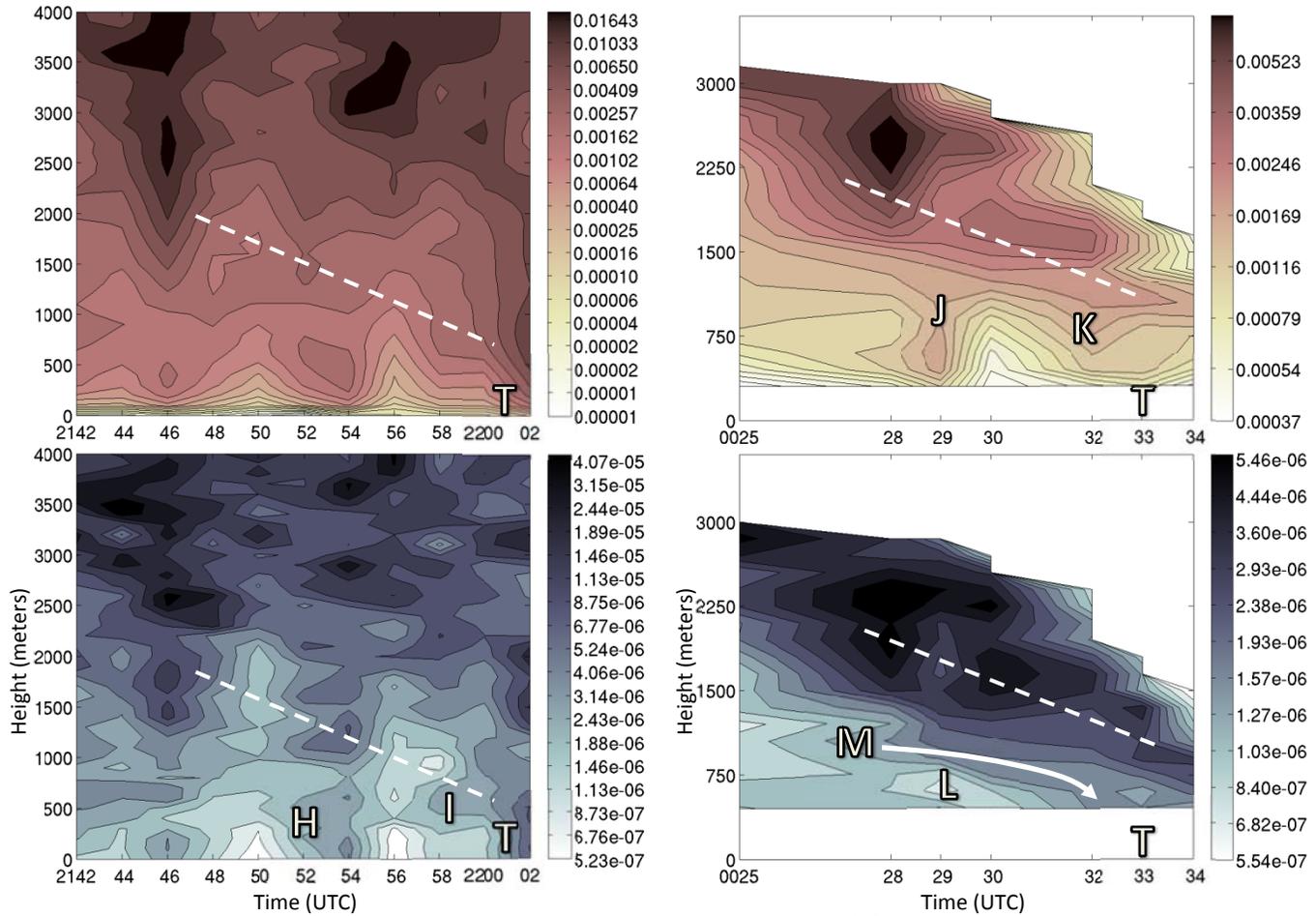

Figure 4. As in Figure 2, but with time-height analysis of peak superhelicity ($S$, m$^{-1}$s$^{-2}$; top row) and vertical gradient of this quantity ($S_z$, m$^{-2}$s$^{-2}$; bottom row) for the Goshen case (left column) and Argonia case (right column). Sloped dashed lines are indicative of descending coherent structure.

## SECTION IV. RESULTS

The peak time-height analyses of the pertinent dynamical quantities generally showed a nearly logarithmic scaling with height. Scaling the data as such revealed surprisingly coherent, propagating dynamical features in the three cases.

### A. Goshen

The evolution of peak vorticity values is examined in Figure 2. A periodic evolution at mid-levels (1.5-4 km) is apparent, with moderately strong values (> .06 s$^{-1}$) extending downward to an altitude of about 1200 meters at 2146 UTC (A), and very strong values (0.2 s$^{-1}$) at about 3500 meters from 2156 to 2200 UTC (B). Strong values (> 0.1 s$^{-1}$) cascade downward from approximately 3000 m at 2154 to near the surface by 2200 (C). Near the surface, a notable increase in vorticity is seen around 2146 UTC (D). Strong surface (0-500m) values are seen from 2152-2156, with some weakening at 2158 UTC, and further strengthening from 2200-2202 UTC (E). This pattern in the peak surface vorticity is consistent with the interpretation given in *MK12* and *K13*. *K13* documented initial tornadogenesis around 2152-2156, with an 'interruption' from 2156-2158, and a resumption of tornadogenesis from 2200-2202 UTC.

There is some correlation with the vorticity pattern and that shown in the time-height maximum total vorticity tendency field (TOT; Figure 3). Periodic peaks in TOT are seen near the surface at 2146, 2152-2156, and 2200-2202 UTC. These surface peaks are accompanied by (but somewhat vertically disconnected from mid-level maxima that are also periodic. The maximum at 2146 extends downward from altitudes well above 3 km to below 1.5 km, while the mid-level maxima at 2152 extends (albeit more dispersed) from well above 3 km down to (more strongly) below 1 km. A striking minimum appears from 2156-2158 UTC at 500-1000 m altitude. A similar pattern is noted in the LTOT field (Figure 3), but there appears to be stronger vertical continuity from the mid-levels to the surface at 2144-2146, 2154 UTC, and especially at 2200-2202 UTC, during tornadogenesis. Additionally, LTOT shows a broader development of higher values at altitudes around 500-1200 m at 2158-2200 UTC, as opposed to TOT at a similar height and time. Furthermore, a stronger (and earlier) minimum is also apparent near the surface at 2156-2158 UTC (correlated with the 'interruption' phase of the Goshen storm).

Figure 4 shows the evolution of peak superhelicity values (*S*). Near the surface (~400 m), *S* shows periodically growing (in depth and duration) maxima at about the same time periods as the other kinematic quantities (2146, and 2152-54 UTC), accompanied by periodically strengthening minima (lulls) at 2148-2150 and 2156-2158 UTC. Very strong values are seen at altitudes above 1.5 km at 2146, with a broad, but weaker maximum extending from 1.5 km to well above 3 km around 2154-2158. Moderately strong values extend vertically through the depth of the domain during tornadogenesis around 2200-2202 UTC.

As shown in Section III, one would expect strong correlation between $S_z$ and LTOT. Indeed, this is evident when comparing Figure 4, which depicts the vertical gradient of superhelicity ($S_z$), and LTOT in Figure 3. Again, there exists stronger vertical continuity in peak values from the surface to mid-levels at 2146, 2154, and 2202 UTC. Similar lowering and pulsing of the maxima is also present. However, notable differences are seen around 2150-2154, and 2158-2200 UTC (noted by H and I); development of shallow (200-800 m) $S_z$ maxima seem to precede more intense surfaces features by 2-4 minutes, as compared to LTOT.

**B. Argonia**

While the dataset for the Argonia storm lacks the same temporal breadth as Goshen, some similar patterns are still revealed in the analysis. *MQ12* interpreted weak tornadogenesis around 0033 UTC, consistent with the surface maximum seen in vertical vorticity (Figure 2). As in the Goshen case, a few minutes before tornadogenesis, there is a formation of a surface maximum at 0028-0029, followed by a brief 'interruption' at 0030, before intensifying again by 0032 UTC. The initial surface maximum (at 0028-0029 UTC) is also somewhat vertically disconnected from (but positioned underneath) a stronger, mid-level maximum (D) at around 2.5 km., while the 'influence' of the mid-level circulation seems to penetrate downward (E) at the time of mild surface weakening (0030 UTC). It is tempting to argue from the spatiotemporal path (E→F) that part of the mesocyclone circulation penetrated all the way to the surface by 0032 UTC. Additionally, it appears the two surface maxima that develop are vertically disconnected from the mid-level circulation (note minima at locations G), although there is uncertainty in these interpretations, given an increasing paucity of data at the mid-levels during this time period. However, when assimilating this dataset into a numerical simulation, *MQ12* did indeed show that the surface tornado became increasingly disconnected from the mid-level mesocyclone starting around this time (see *MQ12*, Fig. 9). Overall, there exists a strikingly similar pattern between the entire Argonia vorticity plot, and a similar time and height subsection of the Goshen plot, outlined as a thin black box in Figure 2.

Figure 3 also shows a similar pattern to Goshen, especially in the LTOT field, where a periodic, descending maxima appears to propagate from 2-3 km at 0025 to the near the surface by 0033 UTC. A small surface maximum appears in TOT at 0029, while a more distinct surface feature is seen a little earlier (0028-0029 UTC) in LTOT. As in the vorticity field, there is a lull (more distinct in LTOT) at 0030; another lull appears at 0034, shortly before the tornado dissipated [*MQ12*]. Similarities also exist with the vertical disconnection between the initial surface maximum and a vertically superpositioned mid-level maximum at 0029, but a deeper vertical continuity (and stronger intensity) appearing at 0032-0033 UTC.

The *S* and $S_z$ fields show a downward-sloped trend in time (Figure 4), although the periodic nature of the superhelicity maxima is more defined at lower altitudes, while $S_z$ is more pulse-like above 1.5 km altitude. Again, a slight vertical disconnect exists between the near-surface and mid-level maxima at 0028-0029 UTC (J), while the surface feature at 0032-0033 UTC (K) appears connected to a larger descending coherent structure. Since the small local maximum of *S* at 0029 UTC is centered at an altitude of around 500 meters, it is not surprising that a minimum in $S_z$ appears at about 600-700 meters at that time (L). There is also a similar pattern to Goshen, where a shallow maximum in $S_z$ is evident just above the surface (about 1000 m) about 4-5 minutes before tornadogenesis, with the feature sloping downward to the surface around the time of tornadogenesis (M, with curved white arrow).

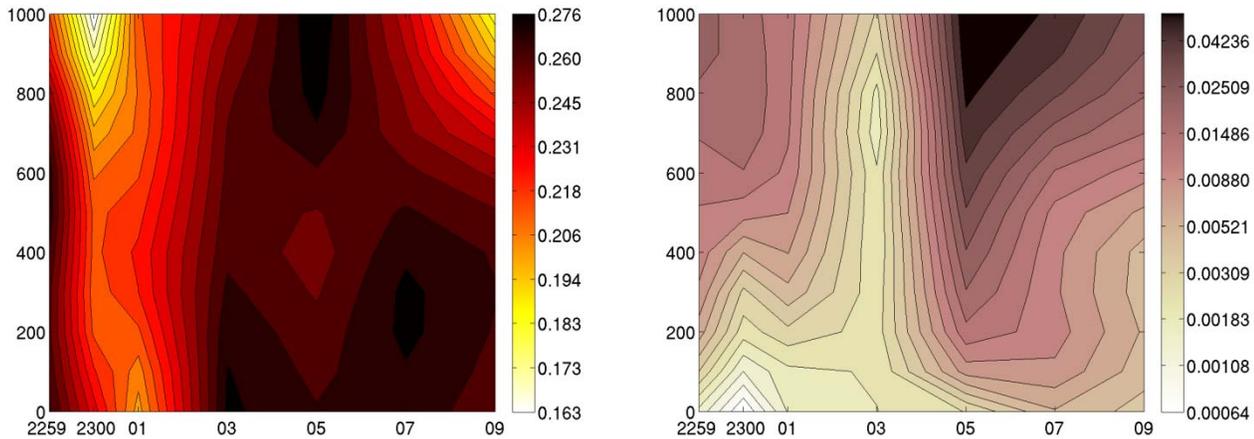

Figure 5.  As in Figure 2, but with time-height analysis of peak vertical vorticity (left) and superhelicity (right) values for the Orleans case.

## C. Orleans

Since this dataset was limited in height (0-1 km) and time (10 minutes), it is difficult to ascertain much information about descending coherent structure in the storm.  However, there are some key features that mirror the other two cases.  The near-surface vorticity field (Figure 5) shows the same pattern (as the other two cases) of a lull (2300-2301 UTC) shortly before renewed intensification (2303 UTC).  This is consistent with the DOW2 and DOW3 peak vorticity pattern reported in *W10*; however, given the slight surface weakening at 2305 UTC, this analysis seems more consistent with the DOW2 reported vorticity (see Figure 7 in *W10*).  This, combined with the 0-800m vorticity maximum appearing in the 0-1 km dataset at 2259 (C), resembles the same pattern of a vertically isolated surface vorticity maximum that precedes a lull, followed by stronger tornadogenesis, as in the other two cases.  Interestingly, the lull in the vorticity field at 2300-2301 actually *precedes* a large minimum seen in all other diagnostic variables related to vorticity tendency (not shown).  However, there may be periodic modes present that are not evident, given the small spatial and temporal sample of the data.  There is also evidence of periodic descent in the superhelicity peak parameter (Figure 5).

## D. Cross-Correlations between Diagnostic Variables

While the figures show strong similarity between salient fields (e.g. LTOT and $S_z$), it is also beneficial to look at the cross-correlations between peak values of several key parameters proximal to the mesocyclone.  There are a fair number of variables that are highly correlated (or anti-correlated, in the case of maximum downward velocity, $w$-) with each other (Table 2).  The top correlated pair, LTOT and TOT  (0.92), is not surprising, given that one describes local maxima of the other.  The next highest correlation magnitude was $w$- and $\omega_h$ (0.91), which is also not expected, given downdrafts are usually considered the primary source for $\omega_h$.  Coming in third, fourth, and fifth places are the previously stated relationships of $S_z$ and LTOT (0.88); $S$ and $\omega_h$ (.85); and finally, $S_z$ and TOT (.81).  Correlations involving UDZ were comparatively weak in most cases.

| Variable | $\omega_h$ | S | $S_z$ | TOT | $\zeta$ | LTOT | UDZ |
|---|---|---|---|---|---|---|---|
| w- | **-0.885** | -0.751 | -0.617 | -0.617 | -0.486 | -0.573 | -0.617 |
|    | **-0.853** | -0.665 | -0.666 | -0.733 | -0.145 | -0.718 | -0.428 |
|    | **-0.987** | -0.881 | -0.510 | -0.631 | 0.026  | -0.778 | -0.599 |
| $\omega_h$ |  | 0.817 | 0.707 | 0.738 | 0.431  | 0.720 | 0.711 |
|    |  | 0.827 | 0.810 | 0.761 | 0.262  | 0.846 | 0.349 |
|    |  | 0.909 | 0.542 | 0.654 | -0.044 | 0.772 | 0.605 |
| S |  |  | 0.705 | 0.714 | 0.506 | 0.704 | 0.651 |
|   |  |  | 0.786 | 0.640 | 0.433 | 0.793 | 0.145 |
|   |  |  | 0.739 | 0.820 | 0.067 | 0.874 | 0.605 |
| $S_z$ | All cross-correlation values: |  |  | **0.865** | 0.257 | **0.876** | 0.597 |
|   |  |  |  | **0.671** | 0.468 | **0.875** | 0.390 |
|   |  |  |  | **0.903** | 0.107 | **0.892** | 0.656 |
| TOT | 1-Goshen |  |  |  | 0.279 | **0.954** | 0.665 |
|   | 2-Argonia |  |  |  | 0.359 | **0.862** | 0.264 |
|   | 3-Orleans |  |  |  | 0.153 | **0.952** | 0.749 |
| $\zeta$ |  |  |  |  |  | 0.220 | 0.273 |
|   |  |  |  |  |  | 0.460 | 0.439 |
|   |  |  |  |  |  | 0.016 | -0.001 |
| LTOT |  |  |  |  |  |  | 0.665 |
|   |  |  |  |  |  |  | 0.307 |
|   |  |  |  |  |  |  | 0.890 |

Table 2. Cross-correlation coefficients between seven kinematic and dynamic variables for the three observed cases.

## SECTION V: DISCUSSION

### A. Descending, pulsing coherent structures

In most of the diagnostic variables, in the three cases, there appeared to be a measure of coherent periodicity and descent with time. Intensification of surface vorticity coincided with the approach of these structures to the surface. Superhelicity (and its vertical gradient) was one of the clearest indicators of this process. Since $S$ and $S_z$ depend on gradients in horizontal vorticity, the next logical step is to question what causes the descending, periodic nature of the horizontal vorticity. The obvious answer is some kind of pulsing, flanking downdraft. This would be consistent with a number of studies showing the correlation of a secondary downdraft pulse and associated secondary rear-flank gust front (SRFGF) coinciding with tornadogenesis (*e.g.* Marquis, 2008; Lee et al., 2012). Our understanding of how the SRFGF acts to intensify the surface vorticity, and the cause of the periodicity of the downdraft strength, is still incomplete. One possibility is that the updraft surges due to changes in buoyancy, which could act to interrupt the downdraft. The regularity of the periodicity would seem to argue against such a random forcing as unsteady convection. Another possible mechanism is a vertically-propagating wave (or wave-like) feature. If this is the case, the rate of descent (vertical phase speed) of this wave-like feature is much less than the vertical velocities involved. By inspection of the superhelicity fields, the time-height slope of periodic descent in the coherent is remarkably consistent: all seem to show a vertical phase velocity of about 1-2 m s$^{-1}$ (*e.g.*, sloped dashed lines in Figure 4). This is much slower than typical downdraft velocities (usually on the order of 10 ms$^{-1}$ or greater). One type of vertically propagating wave that fits this profile would be a vortex Kelvin Wave (*e.g.* Rossi, 2000; Fabre, 2006), traveling axially and/or azimuthally along the periphery of the mesocyclone circulation. To a first approximation, axial phase speeds of these waves are on the order of the rotational velocity, with some dependence on axial wavelength. As these apparent phase velocities are relatively slow, it is likely that either the axial wavenumbers are either large (small wavelengths), or the waves are propagating through the helical updraft (reducing the ground-relative axial phase speed). In either case, this type of wave would lead to descending pulses with alternating radial and vertical velocity anomalies as wave crests pass through a given plane. This would also be manifest as vacillating regions of horizontal vorticity on the periphery of the mesocyclone circulation. These would be diagnosed as coherent, descending superhelicity dipoles. In future work, we will be looking for evidence of these waves in supercells and vortex intensification.

## B. Vorticity intensification mechanisms

Superhelical vorticity enhancement in the mesocyclone circulation (Fig.1) would be consistent with the Kelvin wave mechanism. Since Kelvin waves can only propagate along a vortex in approximate cyclostrophic balance, these waves cannot (initially) reach the surface, as the vortex lines near the base of the mesocyclone circulation spread outward horizontally as the environmental shear. However, localized perturbations in the vorticity field can extend the mesocyclone downward through Kelvin-Helmholtz instabilities, rolling up the vortex sheets associated with the environmental vorticity. Additionally, Kelvin waves can initiate precipitation-enhanced downdraft pulses that can reach the surface, creating a deformation field that can form a sheet of vertical vorticity (*Trevorrow, et al.*, pre-publication). This sheet can then be subject to further local intensification through barotropic instability. This would explain the vertical disconnection of coherent structure in the kinematic fields at around 500-1200 m, near the base of the mesocyclone circulation. Of further note is that any superhelical interaction (Kelvin wave or otherwise) in the vicinity of the vortex sheet can act as localized vorticity perturbations to release barotropic instability.

## C. Operational Considerations

Our findings indicate that significant spatiotemporal coherency in miso-scale regions of vorticity intensification potential can be revealed and tracked by focusing on a) local maxima of various parameters, such as the vortex stretching and tilting terms; b) superhelicity related parameters, given that they are closely tied to the interaction between local maxima of vertical motion (updraft and downdraft pulses) and the mesocyclone; and c) using a logarithmic scale to track coherent features as the propagate downward from well above the boundary layer. Coherency was found even when the magnitude of the values was quite small near the surface compared to 2-3 kilometers aloft. Relying on intrinsically three-dimensional parameters (e.g. $S$ and $S_z$) presents a challenge, as operational forecasters are limited to single-Doppler data. However, improving and simplifying the paradigm of vortex-vortex interactions in supercells (and other turbulent environments) can aid in the ability to infer 3D structure from 2D data. For example, a pulsing, descending convergence line, curved around a mesocyclone, would be a good indicator of a superhelical structure capable of rapid vorticity intensification.

## REFERENCES

Barnes, S. L., 1978: Oklahoma thunderstorms on 29–30 April 1970. Part I: Morphology of a tornadic storm. *Mon. Wea. Rev.,* **106,** 673–684.

Davies-Jones, R., 2008: Can a descending rain curtain in a supercell instigate tornadogenesis barotropically? *J. Atmos. Sci.,* **65,** 2469-2497.

Davies-Jones, R. et al., 2006: Tornadogenesis in supercell storms: What we know and what we don't know. *Preprints, Symp. on the Challenges of Severe Convective Storms, Atlanta, GA, Amer. Meteor. Soc.*

Davies-Jones, R. Trapp, J. and Bluestein, H, 2001: Tornadoes and Tornadic Storms. *Meteorological Monographs*, **28**, 167–222.

Dowell, D., Y. Richardson, and J. Wurman, 2002: Observations of the formation of low-level rotation. The 5 June 2001 Sumner County, Kansas tornado. Preprints, *21st Conf. on Severe Local Storms*, San Antonio, TX, Amer. Meteor. Soc., 465-468.

Fabre, D., Sipp, D.,and Jacquin, L., 2006: Kelvin waves and the singular modes of the Lamb-Oseen vortex. *J. Fluid Mech.*, **551**, 235-274

Fernández-Guasti, M., 2012: Green's Second Identity for Vector Fields. *ISRN Mathematical Physics*, **2012**, 973968.

Fujita, TT 1975: New evidence from the April 3–4, 1974 tornadoes. Preprints, *Ninth Conf. on Severe Local Storms,* Norman, OK, Amer. Meteor. Soc., 248–255.

Galanti, B, and Tsinober, A., 2006: Physical space properties of helicity in quasi-homogeneous forced turbulence. *Phys. Let. A*, **352** , 141-149.

Hide, R., 1989: Superhelicity, Helicity And Potential Vorticity. *Geophysical And Astrophysical Fluid Dynamics*, **48**, 69-79.

Hide, R., 2002: Helicity, superhelicity and weighted relative potential vorticity: Useful diagnostic pseudoscalars? *Q. J. Roy. Met. Soc. A*, **128**, 1759-1762

Jacobitz, F. G. *et al*., 2011: Influence of initial mean helicity on homogeneous turbulent shear flow. *Phys. Rev. E* , **84**, 5, 056319.

Kanak, K. M. and D. K. Lilly, 1996: The linear stability and structure of convection in a mean circular shear. *J. Atmos. Sci.*, **53**, 2578-2593.

Kosiba, K.; Wurman, J.; Richardson, Y.; et al., 2013: Genesis of the Goshen County, Wyoming, Tornado on 5 June 2009 during VORTEX2. *Mon. Wea. Rev.,* **141,** 1157-1181.

Lee, B. D.; Finley, C.A.; Karstens, C. D., 2012: The Bowdle, South Dakota, Cyclic Tornadic Supercell of 22 May 2010: Surface Analysis of Rear-Flank Downdraft Evolution and Multiple Internal Surges. *Mon. Wea. Rev.,* **140,** 3419-3441

Lemon, L. R., and C. A. Doswell, 1979: Severe thunderstorm evolution and mesocyclone structure as related to tornadogenesis. *Mon. Wea. Rev.,* **107,** 1184–1197.

Markowski, P., Y. Richardson, J. Marquis, J. Wurman, K. Kosiba, P. Robinson, D. Dowell, E. Rasmussen, and R. Davies-Jones, 2012a: The pretornadic phase of the Goshen County, Wyoming, supercell of 5 June 2009 intercepted by VORTEX2. Part I: Evolution of kinematic and surface thermodynamic fields. *Monthly Weather Review*, **140**, 2887-2915.

Markowski, P., Y. Richardson, J. Marquis, R. Davies-Jones, J. Wurman, K. Kosiba, P. Robinson, E. Rasmussen, and D. Dowell, 2012b: The pretornadic phase of the Goshen County, Wyoming, supercell of 5 June 2009 intercepted by VORTEX2. Part II: Intensification of low-level rotation. Monthly Weather Review, **140**, 2916-2938.

Markowski, Paul M., 2002: Hook Echoes and Rear-Flank Downdrafts: A Review. *Mon Wea. Rev.,* **130**, 852-876.

Markowski, P. M., J. M. Straka, and E. N. Rasmussen , 2003: Tornadogenesis resulting from the transport of circulation by a downdraft: Idealized numerical simulations. *J. Atmos. Sci.,* **60,** 795–823.

Markowski, P. M., J. M. Straka, E. N. Rasmussen, R. P. Davies-Jones, Y. Richardson, and J. Trapp, 2008: Vortex lines within low-level mesocyclones obtained from pseudo-dual-Doppler radar observations. *Monthly Weather Review*, **136**, 3513–3535.

Marquis, J.; Richardson, Y.; Markowski, P.; et al., 2012: Tornado Maintenance Investigated with High-Resolution Dual-Doppler and EnKF Analysis. *Monthly Weather Review,* **140**, 3-27.

Marquis, J.; Richardson, Y.; Wurman, J.; et al., 2008: Single- and Dual-Doppler Analysis of a Tornadic Vortex and Surrounding Storm-Scale Flow in the Crowell, Texas, Supercell of 30 April 2000. *Monthly Weather Review,* **136**, 5017-5043.

Ricca, R.L. , 1994: The effect of torsion on the motion of a helical vortex filament. *J. Fluid Mech.* **273**, 241-259.

Rossi, M., 2000: Of Vortices and Vortical Layers: An Overview. *Vortex Structure and Dynamics: Lecture Notes in Physics,* **555**, 40-123.

Shiqang, F., Zhemin, T., 2001: On the helicity dynamics of severe convective storms. *Advances in Atmospheric Sciences*, **18**, 67-86.

Straka, J. M., E. N. Rasmussen, R. P. Davies-Jones, and P. M. Markowski, 2007: An observational and idealized numerical examination of low-level counter-rotating vortices toward the rear flank of supercells. *Electronic Journal of Severe Storms Meteorology*, **2**(8), 1–22.

Trevorrow, S.,: Numerical investigation of dynamics leading to tornadogenesis in a supercell environment. Pre-publication.

Wurman, J., K. Kosiba, P. Markowski, Y. Richardson, D. Dowell, and P. Robinson, 2010: Finescale single- and dual-Doppler analysis of tornado intensification, maintenance, and dissipation in the Orleans, Nebraska, supercell. *Monthly Weather Review*, **138**, 4439-4455

Wurman, et al., 2010: An overview of the VORTEX2 field campaign. Preprints, *25th Conf. on Severe Local Storms*, Denver, CO, Amer. Meteor. Soc.

Wurman, J. Richardson, Yvette; Alexander, Curtis; et al., 2007: Dual-Doppler analysis of winds and vorticity budget terms near a tornado. *Monthly Weather Review,* **135**, 2392-2405.

Wurman, J., Straka, J. , Rasmussen, et al., 1997: Design and Deployment of a Portable, Pencil-Beam, Pulsed Doppler Radar, *J. Atm. Ocean. Tech.*, **14**, 1502-12.

Wurman, J., and S. Gill, 2000: Fine-Scale Radar Observations of the Dimmitt, Texas Tornado, *Mon. Weather Rev.*, **128**, 2135-2164.

Wurman, J., 1999: Tornadoes Observed by the Doppler On Wheels Radar, *J. of Visualizat.*, **2**, 1.